\documentclass[10pt,preprint]{aastex}
\usepackage{psboxit}

\PScommands

\newcommand{\latin}[1]{{#1}}

\newcommand{\eg}{\latin{e.g.}}

\def\simless{\mathbin{\lower 3pt\hbox
	{$\,\rlap{\raise 5pt\hbox{$\char'074$}}\mathchar"7218\,$}}} % < or of order
\def\simgreat{\mathbin{\lower 3pt\hbox
	{$\,\rlap{\raise 5pt\hbox{$\char'076$}}\mathchar"7218\,$}}} % > or of order

\newcommand{\band}[2]{\ensuremath{^{#1}\!{#2}}}

\newcounter{thefigs}
\newcommand{\fignum}{\arabic{thefigs}}

\newcounter{thetabs}

\newcounter{address}

\shortauthors{Blanton {\it et al.} (2002)}
\shorttitle{What aspects of galaxy environment matter?}

\begin{document}

\title{ What aspects of galaxy environment matter? }

\author{
Michael R. Blanton\altaffilmark{\ref{NYU}}, 
Andreas A. Berlind\altaffilmark{\ref{NYU}}, 
and David W. Hogg\altaffilmark{\ref{NYU}}}

\setcounter{address}{1}
\altaffiltext{\theaddress}{
\stepcounter{address}
Center for Cosmology and Particle Physics, Department of Physics, New
York University, 4 Washington Place, New
York, NY 10003
\label{NYU}}

\begin{abstract}
We determine what aspects of the density field surrounding galaxies
most affect their properties.  For Sloan Digital Sky Survey galaxies,
we measure the group environment, meaning the host group luminosity
and the distance from the group center (hereafter, ``groupocentric
distance''). For comparison, we measure the surrounding density field
on scales ranging from 100 $h^{-1}$ kpc to 10 $h^{-1}$ Mpc.  We use
the relationship between color and group environment to test the null
hypothesis that only the group environment matters, searching for a
residual dependence of properties on the surrounding
density. Generally, red galaxies are slightly more clustered on small
scales ($\sim$ 100--300 $h^{-1}$ kpc) than the null hypothesis
predicts, possibly indicating that substructure within groups has some
importance. At large scales ($> 1$ $h^{-1}$ Mpc), the actual projected
correlation functions of galaxies are biased at less than the 5\%
level with respect to the null hypothesis predictions.  We exclude
strongly the converse null hypothesis, that only the surrounding
density (on any scale) matters.  These results generally encourage the
use of the halo model description of galaxy bias, which models the
galaxy distribution as a function of host halo mass alone.  We compare
these results to proposed galaxy formation scenarios within the Cold
Dark Matter cosmological model.
\end{abstract}


\section{Introduction: galaxy properties and environment }
\label{intro}

Galaxy properties are a strong function of their environment. In
particular, galaxies in dense environments are older, redder, more
concentrated, higher in surface brightness, and more luminous than
galaxies in voids (\citealt{hubble36a, oemler74a, davis76a,
postman84a, dressler80a, santiago92a, hermit96a, zehavi02a,
norberg02a, hogg03b, blanton05b, kauffmann04a, weinmann05a, maller05a,
martinez06a}).  Clearly, part of understanding galaxy formation must
involve understanding why regions with different initial conditions (a
different initial cosmological density field) result in such different
galaxy populations.

Quantitative understanding of this variation of clustering with galaxy
type has improved in the last twenty years or so with the advent of
large extragalactic samples. Among the papers written on this subject,
what the authors mean by ``galaxy type'' has varied considerably. For
some ``type'' has meant a measure of its optical morphological
properties --- is it an elliptical, a dwarf elliptical, a lenticular,
a spiral, or an irregular? Others have preferred measures of galaxy
structure such as size or concentration, or measures of the galaxy
star-formation history, such as color or emission line flux.  A number
of recent papers (\citealt{kauffmann04a, blanton05b, quintero06a})
have shown that star-formation related properties such as color and
emission line flux are directly affected by environment, while
structural properties such as surface brightness and concentration
(which are more closely related to classical morphology) are not.  In
addition, \citet{kauffmann04a} have shown that environment does not
affect measures of very recent star-formation such as emission lines
(the last 10 Myr) over and above its affect on longer time scale
measures such as optical color (the last 1 Gyr).  Whether other
classical morphological measurements, such as spiral arm properties,
are related to environment directly, remains to be seen. In this
paper, we will simplify based on these results to only consider the
relationship between environment and galaxy colors.

Astronomers have not only differed by what they mean by ``galaxy
type,'' they also have differed in how they measure ``environment.''
Usually their choices have been motivated by practicality --- after
all, the ultimate measure of environment, the local mass density
field, is currently observationally inaccessible. In some cases,
astronomers have used distance from the center of a cluster or surface
density within that cluster as a measure of environment
(\citealt{oemler74a, dressler80a, postman84a, gomez03a,
quintero06a}). In other cases, we have used field samples and measured
the number of nearby galaxies relative to the mean density, using
various measures (\citealt{hashimoto98a, hashimoto99a, balogh04a,
hogg04a, kauffmann04a, blanton05b}). Sometimes, we do not measure the
density around individual objects, but instead measure the average
environments of classes of galaxies. For example, we can measure as a
function of galaxy properties the mean environments around galaxies
(\citealt{hogg03b,blanton05b}) or their correlation functions
(\citealt{davis76a, santiago92a, hermit96a, norberg02a,
zehavi04a}). In this paper, we will consider all of these approaches
to understand what description of the density field is most directly
related to galaxy colors.

All of these analyses have found the same qualitative results. The
``later type'' galaxies --- diffuse, low surface brightness, blue
spiral or irregular galaxies --- populate the low density regions. The
``earlier type'' galaxies --- concentrated, high surface brightness,
red elliptical or S0 galaxies --- populate the high density regions.
However, all of the measures of environment are correlated at least
weakly to each other, so one expects them all to show the same
qualitative trends.  If we can determine what aspects of environment
are {\it most directly} related to galaxy properties, it will be an
important clue in determining the way in which environment affects
galaxy formation. Making this determination is the goal of this paper.

We choose here two classes of environmental measurements: ``group
environment'' measurements and ``surrounding density measurements'' on
varying scales.  Here, we repeatedly use the term ``group
environment'' to refer to two parameters regarding the galaxy
environment: the luminosity of the host group or cluster (group
luminosity) and the radius from the center of that group
(groupocentric distance). By ``surrounding density'' we mean the
density with respect to the mean in a cylinder around each galaxy,
where the cylinder is aligned in the redshift direction to integrate
over nonlinear redshift space distortions. The radius of the cylinder
in the transverse direction determines the scale. We explain these
density measurements in more detail below.

These results yield insights into how the spatial distribution of
galaxies is related to that of the dark matter.  In particular, a
recently developed way of describing the relationship between galaxies
and mass is by using the ``halo occupation distribution'' (HOD), which
quantifies the distribution of galaxies as a function of host dark
matter halo mass (see \citealt{berlind02a} and references therein).
The HOD description is an extremely convenient way of parameterizing
the physics that relates galaxies and mass, and of marginalizing over
the possible relationships when trying to use galaxy clustering to
constrain cosmology (\citealt{abazajian05a}).  It typically assumes
that the distribution of galaxy luminosities and types depends only on
the mass of the host halo, and not on the larger scale density field.
We test this assumption here by associating galaxy groups in the
observations with dark matter halos, and asking whether indeed the
masses of these groups are the only quantities that are relevant to
galaxy properties.

In Section \ref{sdss} we describe the Sloan Digital Sky Survey (SDSS;
\citealt{york00a}) and the \citet{berlind06a} group catalog resulting
from it.  In Section \ref{environs}, we find that both group
luminosity and groupocentric distance are independently related to
galaxy properties. We then consider whether the mean density around
galaxies is related to galaxy colors over and above the dependence
expected to result just from the dependence on group environment. We
only find a large residual dependence on very small scales ($\sim 300$
$h^{-1}$ kpc). Meanwhile, even in regions with very different
densities on large scales, the galaxy population is similar as long as
the group environment is similar. In Section \ref{correlation} we
demonstrate the same results in terms of the correlation functions of
red and blue galaxies.  In Section \ref{others} we compare our results
to similar studies of others, finding agreement. In Section
\ref{theory}, we compare our results to similar investigations in the
theoretical realm. In Section \ref{conclusions} we discuss the
implications of our results for theories of galaxy formation and for
interpreting large-scale structure statistics.

\section{The SDSS spectroscopic sample of galaxies}
\label{sdss}

In order to investigate the questions posed above, we use the SDSS
spectroscopic sample. It is a large, homogeneously selected sample of
galaxies in the local Universe, and is ideal for studying the
relationship between galaxy properties and environment. 

The SDSS is taking $ugriz$ CCD imaging of $10^4~\mathrm{deg^2}$ of the
Northern Galactic sky, and, from that imaging, selecting $10^6$
targets for spectroscopy, most of them galaxies with
$r<17.77~\mathrm{mag}$ \citep[\eg,][]{gunn98a,york00a,abazajian03a}.
Automated software performs all of the data processing: astrometry
\citep{pier03a}; source identification, deblending and photometry
\citep{lupton01a}; photometricity determination \citep{hogg01a};
calibration \citep{fukugita96a,smith02a}; spectroscopic target
selection \citep{eisenstein01a,strauss02a,richards02a}; spectroscopic
fiber placement \citep{blanton03a}; and spectroscopic data reduction.
Descriptions of these pipelines also exist in \citet{stoughton02a}.
An automated pipeline called {\tt idlspec2d} measures the redshifts
and classifies the reduced spectra (Schlegel et al., in preparation).

The spectroscopy has small incompletenesses coming primarily from (1)
galaxies missed because of mechanical spectrograph constraints
\citep[6~percent;][]{blanton03a}, which leads to a slight
under-representation of high-density regions, and (2) spectra in which
the redshift is either incorrect or impossible to determine
($<1$~percent).  In this context, we note that the mechanical
constraints are due to the fact that fibers cannot be placed more
closely than 55$''$; when two or more galaxies have a separation
smaller than this distance, one member is chosen independent of its
magnitude or surface brightness. Thus, this incompleteness does not
bias the sample with respect to luminosity. In addition, there are
some galaxies ($\sim 1$~percent) blotted out by bright Galactic stars,
but this incompleteness should be uncorrelated with galaxy properties.

For the purposes of computing large-scale structure and galaxy
property statistics, we have assembled a subsample of SDSS galaxies
known as the NYU Value Added Galaxy Catalog (NYU-VAGC;
\citealt{blanton05a}).  For most of this paper we use the group
catalog described in \cite{berlind06a}, based on a volume-limited
sample of galaxies from the NYU-VAGC complete down to
$M_{\band{0.1}{r}} - 5\log_{10} h<-19$ within the redshift range
$0.015 < z < 0.068$. For additional tests we will use two alternate
group catalogs complete to $M_{\band{0.1}{r}} - 5\log_{10} h<-18$ and
to $M_{\band{0.1}{r}} - 5\log_{10} h<-20$, and spanning redshift
ranges of $0.015 < z< 0.045$ and $0.015 < z < 0.10$ respectively.
\citet{berlind06a} identify these groups using a friends-of-friends
algorithm (see e.g., \citealt{geller83a, davis85a}) with perpendicular
and line-of-sight linking lengths equal to 0.14 and 0.75 times the
mean inter-galaxy separation, respectively.  Mock galaxy catalogs
demonstrate that these parameters produce galaxy groups that most
closely resemble the underlying dark matter halos. We have included
groups with $N_{\mathrm{gals}}=1$ or 2 (singles and pairs) for some of
our investigations here, but for most test simply use those with
$N_{\mathrm{gals}} \ge 3$.

We define the group luminosity to be the sum of the luminosities of
all the galaxies in the group that appear in the volume-limited
catalog (in this case those with $M_{\band{0.1}{r}} - 5\log_{10}
h<-19$).  We calculate rough mass estimates for the clusters using the
group luminosity function and assuming a monotonic relation between a
group's luminosity and the mass of its underlying dark matter halo.
By matching the measured space density of clusters to the theoretical
space density of dark matter halos (given the concordance cosmological
model and a standard halo mass function), we assign a virial halo mass
to each cluster luminosity.  This determination ignores the scatter in
mass at fixed cluster luminosity and is only meant to yield rough
estimates.  By these estimates, the minimum group mass than can host
any galaxy in our sample is about $5\times 10^{11}$ $h^{-1}$
$M_\odot$.

Each group has an associated ``virial radius'' related to this mass of
\begin{equation}
r_{\mathrm{vir}} =
\left(\frac{3}{4\,\pi}\frac{M}{200\,\rho_{o}}\right)^{\frac{1}{3}}
\quad ,
\end{equation}
where $M$ is the estimated mass of the cluster and $\rho_{o}$ is the
current mean density of the Universe.  We use the group centers given
by \citet{berlind06a}, which are the mean of the member galaxy
positions.  We define the groupocentric distance $r_P$ as the
projected distance measurement from each galaxy to the center of its
host group. We then scale each groupocentric distance by
$r_{\mathrm{vir}}$. We have also tested the effect of using a group
center defined by the densest location in the group on 300 $h^{-1}$
kpc scales, finding no significant difference in our conclusions.

In this study, we will also also the surrounding densities of each
galaxy. We do so by counting galaxies in annuli extended 2000 km
$s^{-1}$ long in the redshift direction, centered on the galaxy of
interest. The five annuli we use are: $0.01 < r_T < 0.1$, $0.1 < r_T <
0.3$, $0.3 < r_T < 1$, $1 < r_T < 6$, and $6 < r_T < 10$ $h^{-1}$ Mpc.
We use random catalogs to estimate how many galaxies we expect in each
cylinder, accounting for holes in the survey and the edges of the
sample (described using the geometry information in the
NYU-VAGC). The ratio of these two is then $N/N_{\mathrm{exp}} =
1+\delta$, where each annulus above we denote $\delta_{0.1}$,
$\delta_{0.3}$, $\delta_{1}$, $\delta_{6}$, and $\delta_{10}$
respectively.

\section{Galaxy colors and environment}
\label{environs}

\subsection{Color correlates with group luminosity}

\begin{figure}
\figurenum{\fignum}
\plotone{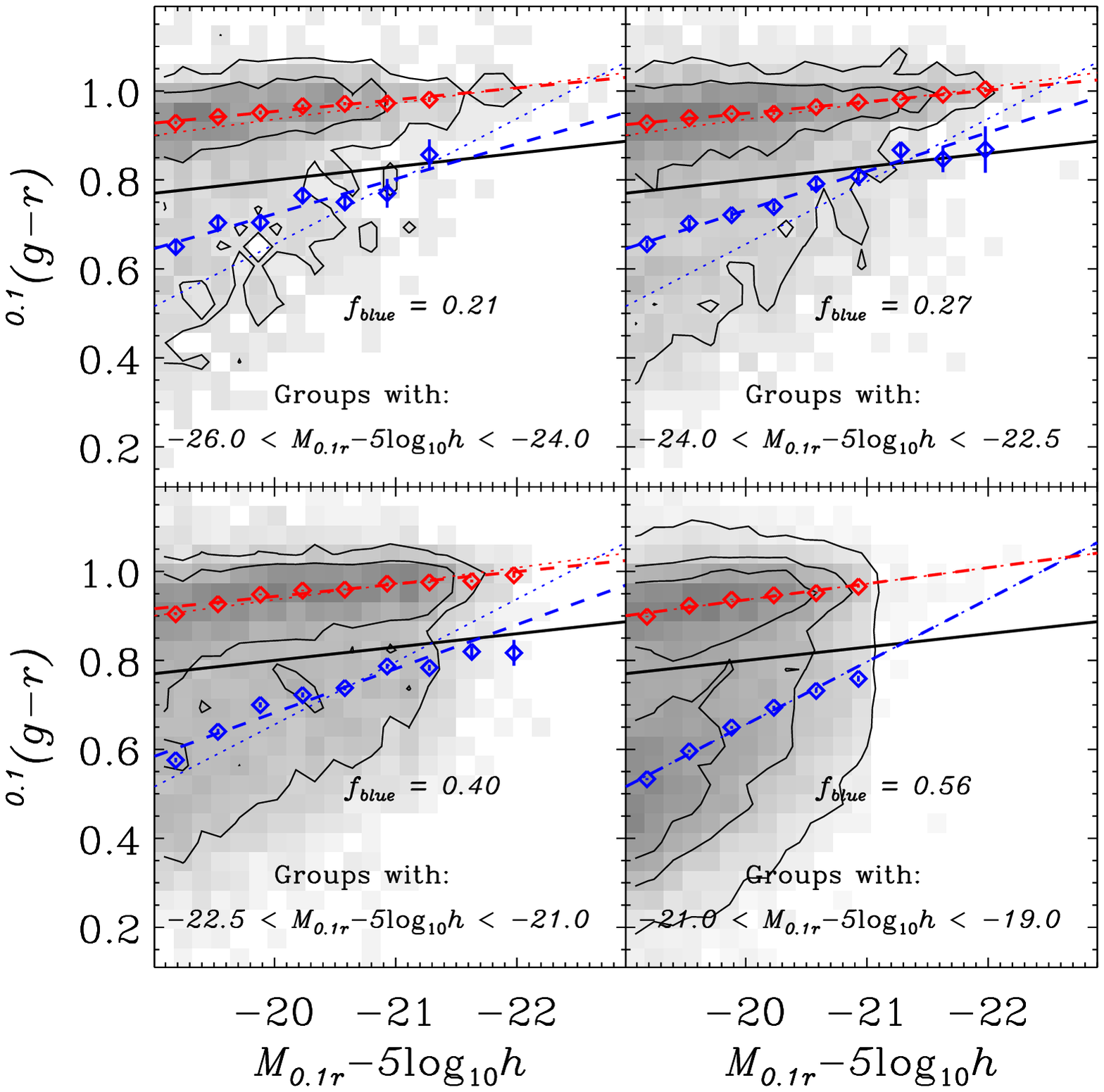}
\caption{\label{cmd_gabsmr} Group luminosity affects the
color-magnitude distribution of galaxies. In each panel, grey scale is
related to number density of galaxies with a given color and absolute
magnitude. The solid line is the same in each panel and represents our
division between ``red'' and ``blue.'' The fraction
$f_{\mathrm{blue}}$ we consider blue by this criterion is listed in
each panel. The dashed lines and diamonds are the fits to the red and
blue sequences described in the text (the dotted lines are identical
in each panel and equal to the dashed lines in the lowest luminosity
set of groups).  Each panel corrsesponds to groups of the listed range
of total absolute magnitudes. As one goes from smaller to larger
groups, the blue fraction decreases, though the positions of the red
and blue sequences do not change much. Note that the cut-off seen on
the right-hand side of the panels (most prominently in the $-21.0 <
M_{\band{0.1}{r}} - 5 \log_{10} h < -19.8$ panel in the lower right)
is imposed by the lower limit on the absolute magnitude in each
panel.}
\end{figure}

We find that galaxy colors are a strong function of their group
luminosity.  Figure \ref{cmd_gabsmr} shows the color-magnitude diagram
of galaxies in our catalog in several bins of group
$\band{0.1}{r}$-band absolute magnitude. We use a definition of
``blue'' and ``red'' according to whether a galaxy is bluer or redder
than the solid line in each panel, which follows the blue edge of the
red sequence using the formula:
\begin{equation}
\label{colorcut}
\band{0.1}{(g-r)}_c = 0.80 - 0.03 (M_{\band{0.1}{r}} + 20.).
\end{equation}
According to this definition, the blue fraction (listed in each panel)
decreases with increasing group luminosity. 

In addition to the blue fraction we can fit the blue and red sequence
positions as a function of luminosity. We do so as follows: in bins of
luminosity we fit the color distribution with a mixture of two
gaussians, in the manner of \citet{baldry04a}. The top row of diamonds
in each panel of Figure \ref{cmd_gabsmr} shows the mode of the redder
gaussian of each fit, the bottom row shows the mode of the bluer
gaussian. In the smallest groups, the mode of the bluer gaussian has a
clear significance, since it corresponds to a mode of the full
distribution.  In the largest groups, the significance of the blue
mode is less clear, though we still include the second gaussian in the
fit to model the blue tail of galaxies in the large groups.  The
dashed lines in each panel show a linear regression on absolute
magnitude for each sequence independently. The dotted lines in each
panel are all identical and show the fit for the lowest luminosity
groups.

For the blue galaxies, there is a change in the slope of the blue
sequence such that the highest luminosity blue galaxies are fixed in
color as a function of environment, but the lower luminosity blue
galaxies become redder in large groups by about 0.1 mag. For the red
galaxies, the red sequence is almost fixed.  There is a trend of about
0.02 mag in color across this range of environment, which is
statistically significant, but we believe verifying this trend
requires a better understanding of the galaxy photometry in SDSS than
currently exists. For example, in rich clusters the colors of the
smaller galaxies may be contaminated by overlapping larger galaxies
(\citealt{masjedi06a}).  Without a more careful examination than space
affords here, we do not make much of this trend. \citet{hogg04a} and
\citet{balogh04b} have previously noted these trends in very similar
data sets.

Thus, galaxy colors are a strong function of their group luminosity,
with the most dramatic change being the reduction of the blue
fraction, and more minor changes occurring in the actual location of
the sequences.  We also note that the reduction in blue fraction
occurs even for the smallest groups, where ram pressure stripping of
the cold disk gas is highly unlikely to affect galaxies of these
luminosities.

\subsection{Color correlates with groupocentric distance}

\begin{figure}
\figurenum{\fignum}
\plotone{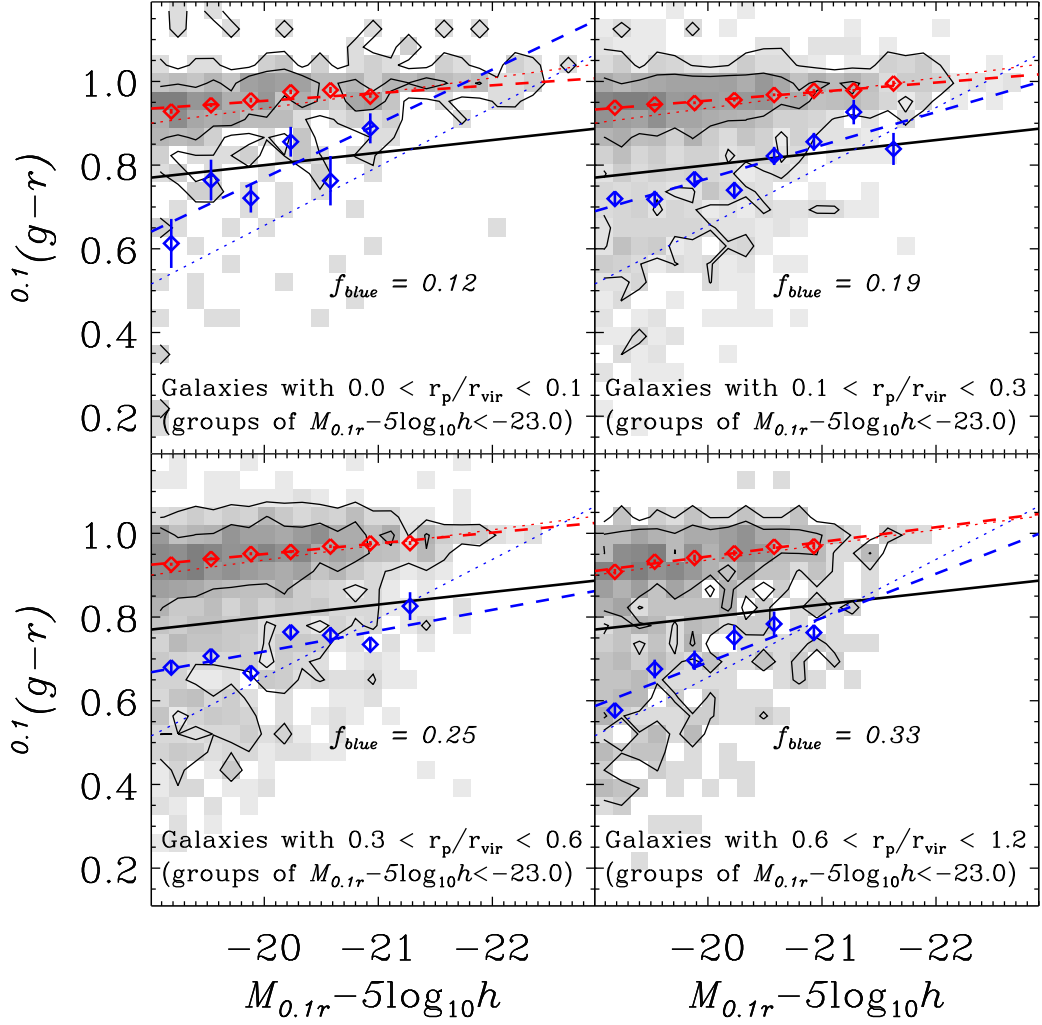}
\caption{\label{cmd_rp} For high luminosity groups, the groupocentric
distance affects the color-magnitude distribution. Similar to Figure
\ref{cmd_gabsmr}, but now concentrating on high luminosity groups
($M_{\band{0.1}{r}} - 5 \log_{10} h < -23$), and dividing galaxies by
the projected distance $r_p$ from the center of the group relative to
the virial radius $r_{\mathrm{vir}}$. The dotted lines here are again
the dotted lines for the low luminosity set of groups.}
\end{figure}

Galaxy colors are also a function of their groupocentric distance.
Figure \ref{cmd_rp} shows the color-magnitude diagram in bins of
$r_p/r_{\mathrm{vir}}$, in groups with total absolute magnitudes
$M_{\band{0.1}{r}} - 5\log_{10} h < -23$.  For such groups, we see a
dependence of galaxy colors on the groupocentric radius. In less
luminous groups there are weaker trends of color with radius, mostly
because single-member groups become important in this regime, and such
galaxies tend to be blue (and are by definition at the centers of
their groups). The dotted lines in this figure are the same as for
Figure \ref{cmd_gabsmr}, the fit to the sequences of the low
luminosity groups. The changes with groupocentric distance of the
positions of the red and blue sequences are similar to the changes
with group luminosity. Thus, at least in luminous groups the trends
with groupocentric distance are similar to the trends with group
luminosity: the blue fraction decreases significantly in the centers
of the groups and the positions of the sequences change slightly.

\begin{figure}
\figurenum{\fignum}
\plotone{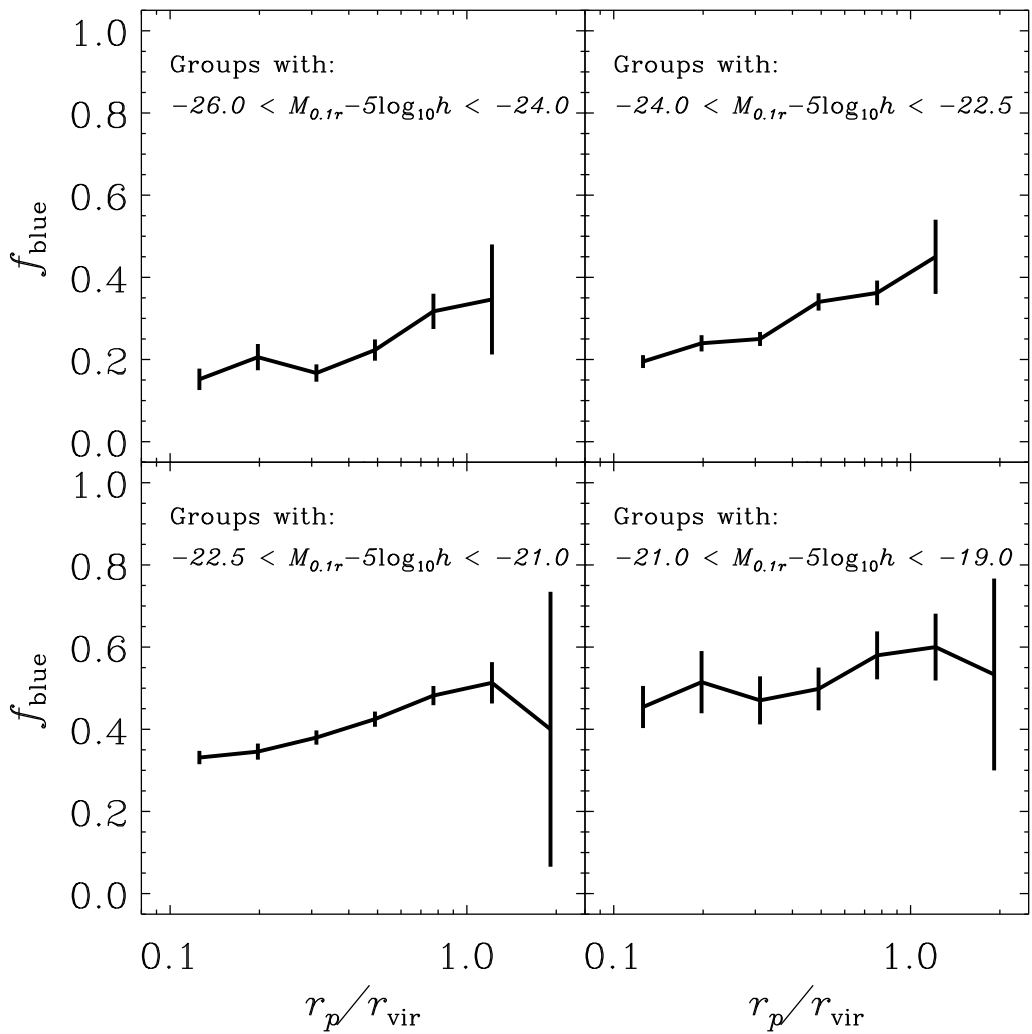}
\caption{\label{bf_rp} The galaxy population is redder at the centers
of groups for all group sizes. Blue fraction as a function of distance
from the center of groups, for four ranges of group luminosity (same
four as used in Figure \ref{cmd_gabsmr}. (The innermost point includes
galaxies all the way to the center of each group).  At all group
luminosities, there is a smaller blue fraction for galaxies near the
centers of the groups. For this figure, we have included only groups
with three or more members (otherwise the distance from the ``center''
is meaningless to consider). }
\end{figure}

Figure \ref{bf_rp} shows the same results, but simply as the blue
fraction as a function of $r_p/r_{\mathrm{vir}}$ for the same ranges
of group luminosities from Figure \ref{cmd_gabsmr}. We here limit the
groups to those with three or more members, since for singles and
pairs the groupocentric radius is close to meaningless.  Thus, both
group luminosity and radius from the center of the group are closely
(and independently) related to galaxy colors, similar to previous
findings (\citealt{dressler80a, whitmore93a, lewis02a, gomez03a,
weinmann05a,quintero06a}).

\subsection{How to test whether only group environment matters }

Neither of the previous subsections have surprising results: they are
consistent with the decades-old understanding of the distribution of
galaxy properties. However, we would like to explore whether the group
environment is the only thing that matters.  Does the surrounding
density field, per se, matter at all?  For example, does the position
of the group in the larger scale density field matter?  In this and
the following sections we will be testing this simple null hypothesis:
can the relationship between galaxy colors and group environment
predict the dependence of galaxy colors on the surrounding density
field at all scales?  

\begin{figure}
\figurenum{\fignum}
\plotone{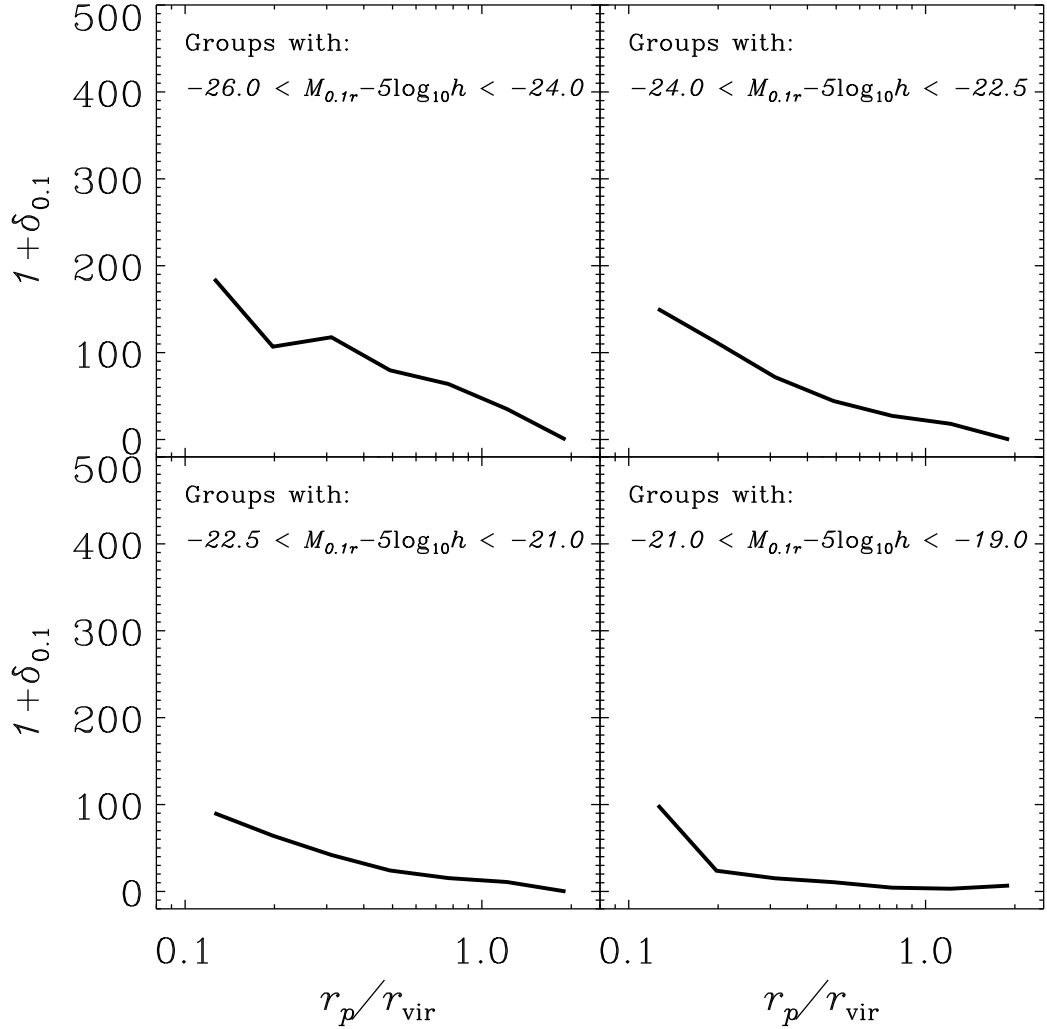}
\caption{\label{den_rp} Density increases towards the centers of
groups. Mean relative density $1+\delta_{0.1}$ on scales of $100$
$h^{-1}$ kpc as a function of distance from the center of groups, for
four ranges of group luminosity (same four as used in Figure
\ref{cmd_gabsmr}.  For this figure, we have included only groups with
three or more members (otherwise the distance from the ``center'' is
meaningless to consider).}
\end{figure}

If the surrounding density really were an important variable, the
residual surrounding density relative to the mean at each group
luminosity and radius would be related to galaxy colors. Consider
Figure \ref{den_rp}, showing the mean density on 100 $h^{-1}$ kpc
scales as a function of radius for groups of various luminosities.
That is, it shows the mean density as a function of group
environment. Obviously, since we know that color correlates with
radius and luminosity, color will correlate with density as well, even
under our null hypothesis.  However, if the surrounding density is
independently important, then a region within a group which is
particularly dense relative to the mean for its group environment
shown in Figure \ref{den_rp} might have a redder (or bluer)
population. Similarly, if the large scale density field is an
important variable, a group surrounded by many other groups would have
a different galaxy population to a similarly sized one that is not.

In the rest of this paper, we compare the actual relationship between
galaxy colors and density field to that predicted by the null
hypothesis.  We construct these predictions by classifying the
environments of galaxies according to their group luminosity and
groupocentric distance, and then shuffling the colors of galaxies with
similar group environments.  This procedure leaves the relationship
between group environment and color intact, and leaves the
relationship between group environment and density intact, but
scrambles the direct relationship between the density field and color.
Then, whatever analysis we perform on the actual galaxy catalogs we
can similarly perform on this shuffled catalog to produce the null
hypothesis prediction.

Note that for small groups (singles and pairs) the group luminosity is
very uncertain, since adding or subtracting a single galaxy from the
group has a large effect. Additionally, the groupocentric distance is
poorly defined in these cases. For these reasons, in many of the tests
below we exclude galaxies in singles and pairs, and concentrate on
galaxies in groups with $N_{\mathrm{gals}} \ge 3$.

\subsection{The large-scale density field matters only a little}

\begin{figure}
\figurenum{\fignum}
\plotone{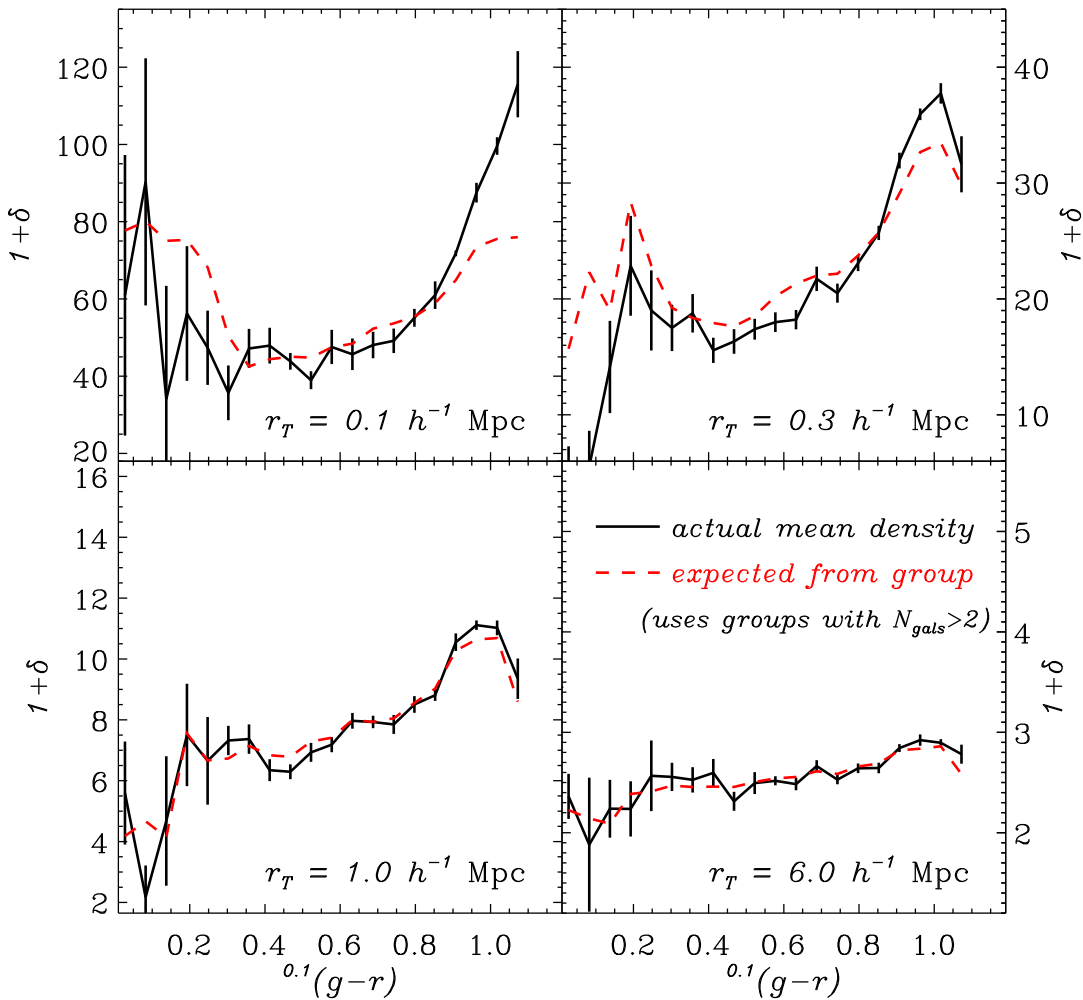}
%\caption{\label{m2_den_color_shuffle} Same as Figure
%	\ref{den_color_shuffle}, but only using galaxies in groups with
%	$N_{\mathrm{gals}} > 2$. The residuals on large scales have largely
%	disappeared.}
\caption{\label{m2_den_color_shuffle} On very small scales ($\sim 300$
$h^{-1}$ kpc), but {\it not} on larger scales, the local density
contains information over and above that predicted by the relationship
between group environment and color.  In each panel, the solid line
shows the mean density as a function of color for galaxies in the
sample, at the scale noted in each panel. We restrict to galaxies in
groups with $N_{\mathrm{gals}} \ge 3$, for which the group
identifications and groupocentric distances are most secure. The
dashed line is the result after shuffling the colors of galaxies with
approximately the same group environment.  On large scales ($r_T> 1$
$h^{-1}$ kpc), the two curves agree well, meaning that galaxy colors
are not affected by the density field (except insofar as different
group environments correspond to different densities). On small scales
($r_T \sim 300$ $h^{-1}$ kpc), on the other hand, the effects can be
as high as 50\%, indicating that the small scale clumping within the
groups may directly relate to galaxy colors.}
\end{figure}

Figure \ref{m2_den_color_shuffle} shows the mean density on various
scales around galaxies as a function of color. The dashed line shows
the expected mean density under the null hypothesis, assuming that the
galaxy population depends only on group environment and not
independently on the density. Here we have restricted the sample to
galaxies in groups with $N_{\mathrm{gals}}\ge 3$, because the
shuffling test is less appropriate for singlets and pairs, whose
groupocentric distances mean very little and whose group luminosities
can be greatly affected by misclassifying group membership.

Consider the scales $r_T \ge 1$ $h^{-1}$ Mpc and greater. On these
scales the solid and dashed lines generally agree better than about
5\%. Thus, the density on these scales is only weakly related to
galaxy colors, once the group environment is known. It is therefore
clear that once we know the group environment, the large-scale
environment (e.g., at 6 $h^{-1}$ Mpc) is not closely related to galaxy
colors.  We have also examined the positions of the blue and red
sequences as a function of the large scale environment, and find only
very small changes in those positions ($<0.03$ mag in color).  From
the point of view of understanding galaxy formation, these results
indicate that the large-scale density field has only a small effect on
galaxy properties.

\subsection{The small-scale density field matters a little bit more}

The small-scale density field is generally less important than the
position of the galaxy within the group. For example, Figure
\ref{den_rp} shows that $\delta_{0.1} \sim 50$ is typical of the
centers of small groups and the outskirts of large groups.  However,
the outskirts of large groups are much redder ($f_{\mathrm{blue}} \sim
0.35$) than the centers of the smallest groups ($f_{\mathrm{blue}}
\sim 0.5$).  Nevertheless, at a fixed group environment, there can be
considerable variation in the small-scale surrounding density field,
because of clumping within the group.  Does this clumping on small
scales matter? As we show here, yes, a little.

Consider the scales $r_T \le 300$ $h^{-1}$ kpc in Figure
\ref{m2_den_color_shuffle}. On these scales it is clear that the
surrounding density is important even given the group environment. The
sense is that very red galaxies tend to be in dense environments on
small scales (relative to the density expected for their groupocentric
distance and a particular group luminosity). Several possible
explanations of this result exist. One is that there is surviving
substructure in the groups from accretion of smaller groups, and that
such substructures have the reddest galaxy populations. A second is
that our group centers (which are associated with the mean positions
of the galaxies) are inappropriate and a more appropriate center would
be the density peak of the group. In fact, we have tested this
possibility by defining the center of the group as the densest
location in the group (instead of the geometric center); our results
do not change significantly with this change in definition, indicating
that this explanation does not account for the effects we see.  A
final possibility is that because our shuffling procedure mixes true
groupocentric radii due to projection effects, some of this extra
information in the density field is because it can be a better
indicator of whether the galaxies are near the true group center.

\subsection{ Blue fraction as a function of environment shows the same}

\begin{figure}
\figurenum{\fignum}
\plotone{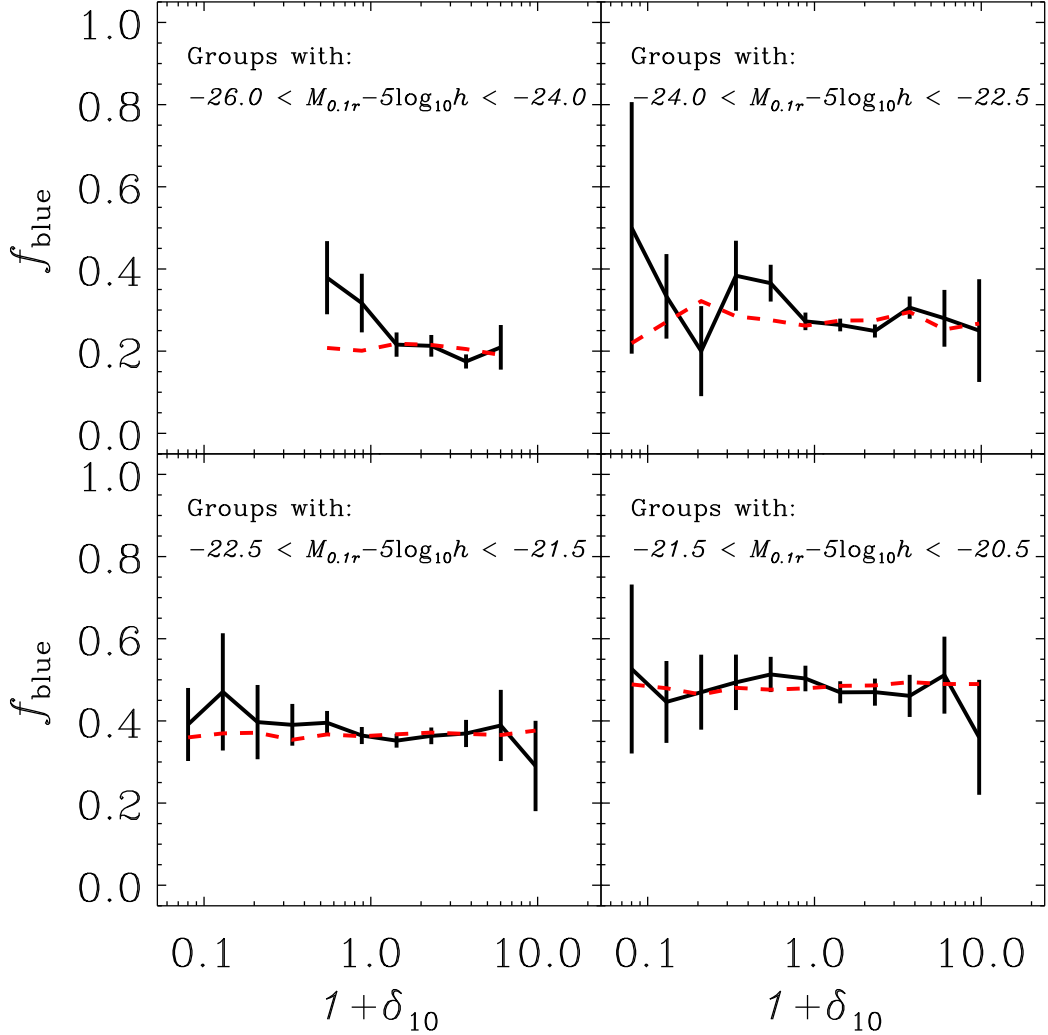}
\caption{\label{bf_den_large_10mpc} Blue fraction as a function of
	large-scale density (in the annulus $6 < r_T < 10$ $h^{-1}$ Mpc),
	for galaxies in groups of various luminosities, for
	$N_{\mathrm{gals}} \ge 3$. The blue fraction is a very weak function
	(at best) of the large-scale density.  The dashed line shows the
	null hypothesis expectation if only the group luminosity and
	groupocentric distance matter.}
\end{figure}

The mean density is a crude measurement --- there could be large
effects that only happen in the densest regions or in the voids that
leave the mean density relatively unperturbed.  To investigate this
possibility, Figure \ref{bf_den_large_10mpc} shows the blue fraction
for groups of different luminosity classes as a function of large
scale density, here calculated around the center of each group for $6
< r_T < 10$ $h^{-1}$ Mpc. The typical number of tracer galaxies in
this annulus is $\sim 50$, suggesting that the errors in the density
estimate are about 15\%, smaller than our bin widths in this
figure. There is very little statistically significant dependence on
density. In fact, we can test explicitly for a difference between what
we would expect by performing the shuffling described above. The
dashed line shows this null hypothesis prediction, which is no trend
--- unsurprisingly, on these large scales.  For the actual galaxies,
the only significant trend is in the most luminous groups, which is
inconsistent with no trend at about 2$\sigma$. This signal is due
mostly to two groups with high blue fractions ($\sim 40\%$) which are
in underdense regions. We note that when using a brighter volume
limited sample (with $M_{\band{0.1}{r}}- 5\log_{10} h < -20$), which
has a larger number of large clusters, we do not find any such trend.

\begin{figure}
\figurenum{\fignum}
\plotone{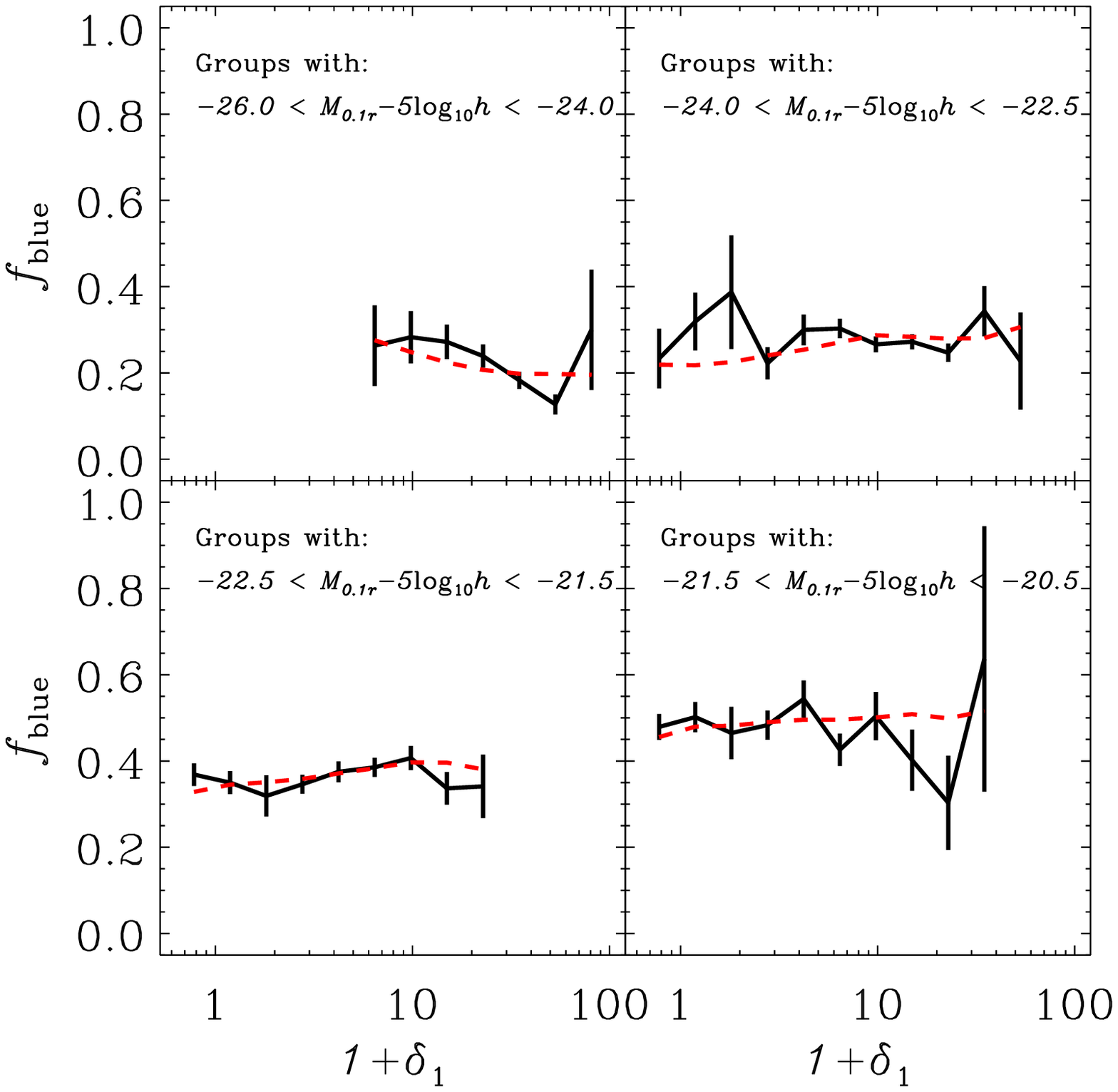}
\caption{\label{bf_den_large_1mpc} Similar to Figure
	\ref{bf_den_large_10mpc}, but for the 
	annulus $0.3 < r_T < 1$ $h^{-1}$ Mpc.}
\end{figure}

\begin{figure}
\figurenum{\fignum}
\plotone{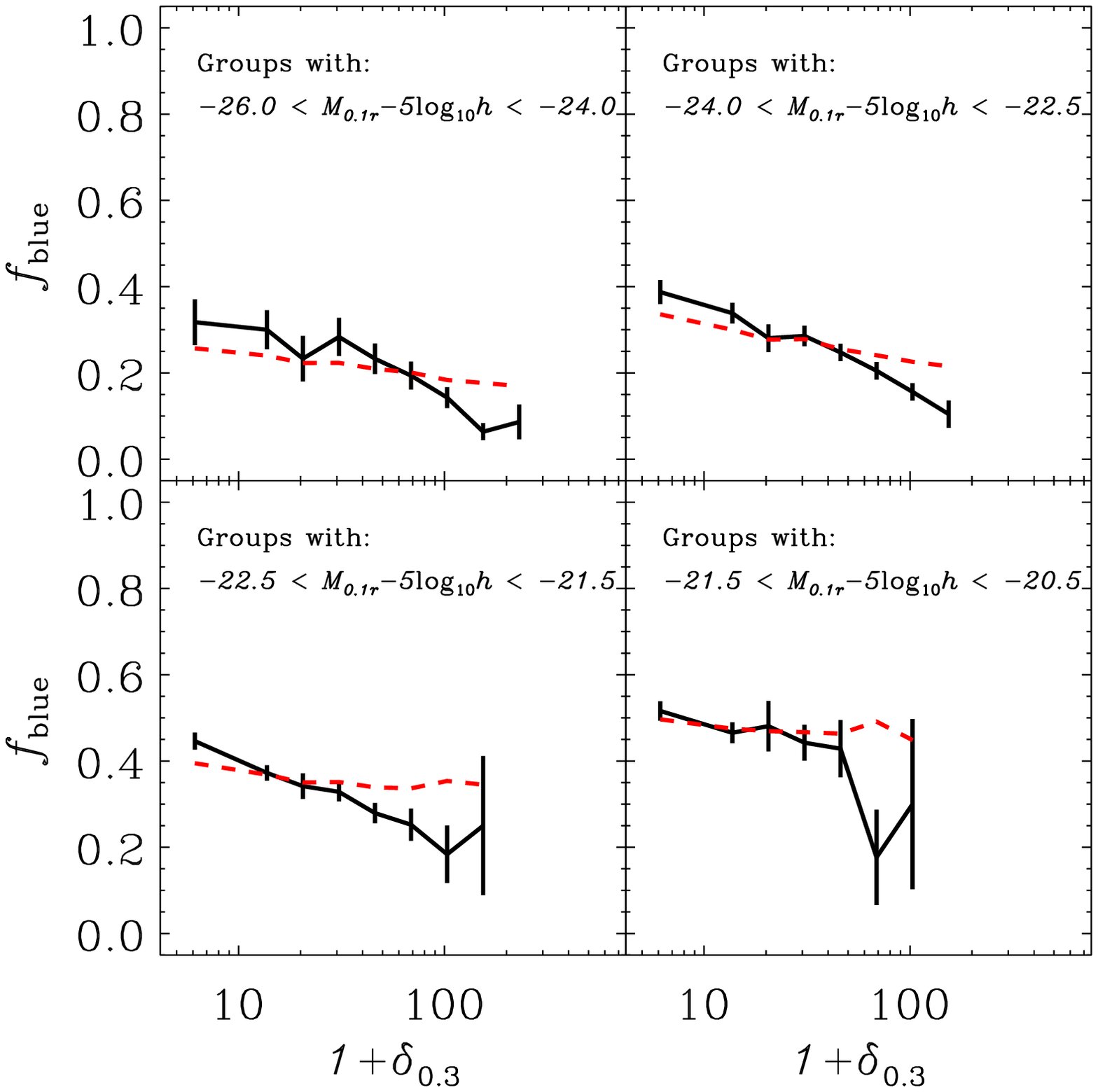}
\caption{\label{bf_den_large_300kpc} Similar to Figure
	\ref{bf_den_large_10mpc}, but for the 
	annulus $0.1 < r_T < 0.3$ $h^{-1}$ Mpc.}
\end{figure}

In Figures \ref{bf_den_large_1mpc} and \ref{bf_den_large_300kpc} we
perform these tests using the smaller scale density estimate
$\delta_1$ and $\delta_{0.3}$. Here the null hypothesis expectation is
no longer flat, since there is a trend of surrounding density with
groupocentric distance at these scales.  At 1 $h^{-1}$ Mpc we find our
results are very close to the null hypothesis expectation, indicating
that the 1 $h^{-1}$ Mpc density field still adds very little
information over and above the group environment.  At 300 $h^{-1}$
kpc, we begin to see a dependence of blue fraction on density that
exceeds the null hypothesis expectation.  This last result is
consistent with what we found in the previous subsection: the smallest
scale densities do matter somewhat over and above the group
environment indicators.

Figures \ref{bf_den_large_10mpc}--\ref{bf_den_large_300kpc} make an
additional point. In each case, at fixed density, the blue fraction is
still a strong function of group luminosity. Thus, on no scale is the
surrounding density a {\it sufficient} description of environment to
explain the trends of galaxy colors with group environment.

\section{Group environment and the correlation function} 
\label{correlation}

How does all of this affect the galaxy correlation function? After
all, as we explain further below, one of the possibilities we want to
test is how adequately we can model the relationship between
environment and colors using the HOD description, and one of the main
uses of that description is to model correlation functions (e.g.,
\citealt{seljak00a, scoccimarro01a, berlind02a, zheng05a}). On the
other hand, if our null hypothesis is incorrect and one needs to know
more than the dependence of galaxy color on group/halo environment in
order to understand the relative correlation functions of galaxies of
different types, then the halo model technique either does not make
much sense or needs to be augmented. In this section, we directly test
the possibility that we need to know more than the group environment.

The correlation function $\xi(r)$ is the excess probability relative
to a Poisson distribution of finding a galaxy a distance $r$ away from
a given galaxy.  Because of peculiar velocities distorting the Hubble
flow in the redshift direction, it is often useful to study this
probability as a function of two variables: $\xi(r_p, \pi)$, where the
separation parallel to the line of sight is $\pi$ and the transverse
separation is $r_p$.  Here we will not be interested in the dependence
on the redshift separation so we will measure a projection of this
function onto the transverse direction (\citealt{davis83a}):
\begin{equation}
w_p(r_p) = 2 \int_0^\infty\, d\pi \xi(r_p, \pi).
\end{equation}
In practice we integrate out to $\pi = 40$ $h^{-1}$ Mpc, following
\citet{zehavi05a}. To estimate $\xi(r_p, \pi)$ we use the \citet{landy93a}
estimator of the correlation function.

\begin{figure}
\figurenum{\fignum}
\plotone{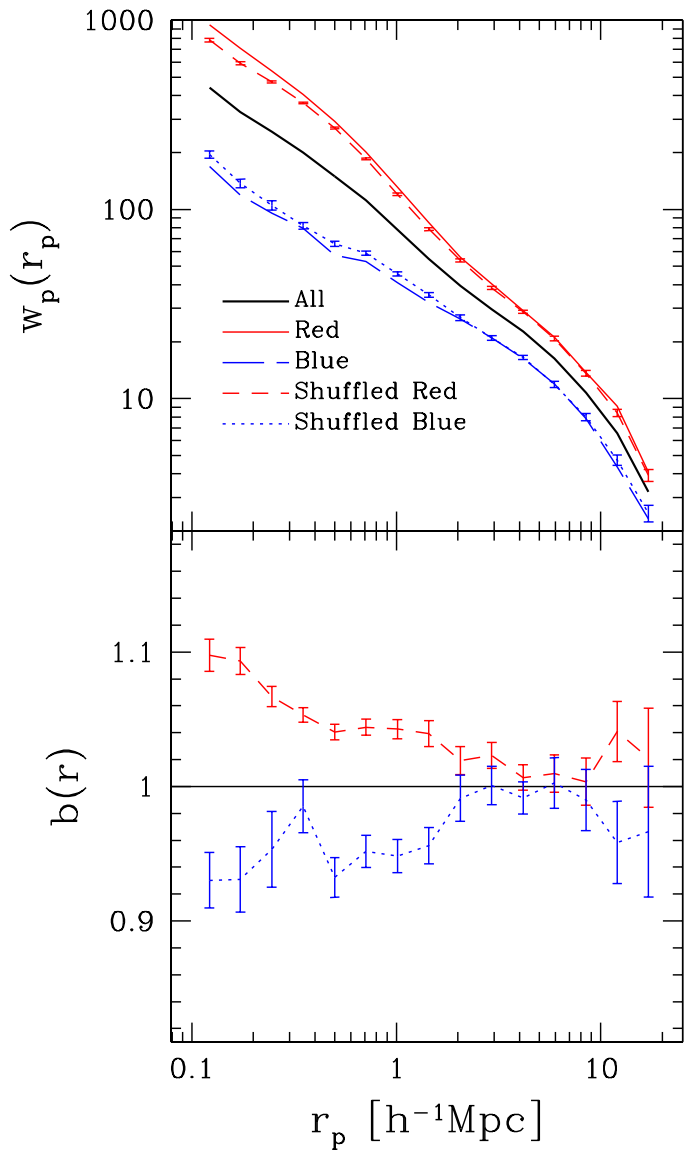}
\caption{\label{wp} Demonstration that group environment explains most
	of the relative bias between galaxy types. {\it Top panel}:
	correlation function of red galaxies (upper lines), blue galaxies
	(lower lines) and all galaxies (thick middle line). The solid line
	for the red galaxies and the long-dashed line for the blue galaxies
	are the actual correlation functions. The short-dashed line for the
	red galaxies and the dotted line for the blue galaxies are the
	correlation functions of the shuffled samples.  The error bars
	indicate the variance among twenty independent shuffles. The
	short-dashed and dotted lines correspond to shuffling the galaxy
	colors among galaxies in groups of similar mass. {\it Bottom panel}:
	dashed and dotted lines are the bias of the original correlation
	function with respect to the shuffled correlation function.}
\end{figure}

In order to ask whether the relative correlation function of red and
blue galaxies is explained by group environments alone, we execute the
same shuffling test we used above, shuffling the colors among galaxies
with similar group environments. Unlike in the previous section, we
include all galaxies in the shuffling, even singles and pairs. Figure
\ref{wp} shows the results. In the top panel, the solid, long-dashed,
and thick solid lines are the actual correlation functions of blue
(lowest), red (highest), and all (middle) galaxies.  The dotted and
short-dashed lines are the null hypothesis predictions of the
correlation functions of blue and red galaxies, resulting when we
shuffle galaxies with similar groupocentric radii and host group
luminosities. In fact, we have shuffled twenty independent times and
here plot the mean of those twenty different shuffles.  The error bars
in Figure \ref{wp} are simply the variance among the twenty
realizations.  The bottom panel shows the square root of the ratios of
the actual correlation functions to the shuffled correlation
functions.

In all cases, the biases are less than 5\% on scales greater than 1
$h^{-1}$ Mpc. These results are consistent with those in the previous
section: the relative strength of the large scale correlation
functions of red and blue galaxies contains very little information
that is not already in the variation of color with group environment.

%We note in passing that the deviations for red and blue galaxies on
%large scales are both decreases in the correlation function.  This
%decrease occurs because the phases of the two density fields are {\it
%slightly} different even on these large scales. That is, the
%cross-correlation of the red and blue galaxies is slightly lower than
%the geometric mean of the autocorrelation functions. The correlation
%coefficient, the ratio $r$ of the cross-correlation to the geometric
%mean, is still greater than $0.9$, consistent with previous estimates
%using similar (\citealt{zehavi05a}) and other techniques
%(\citealt{blanton00a, wild05a}).

At smaller scales, the shuffled correlation functions begin to differ
somewhat from the actual ones by small amounts, with up to a 10\% bias
on the smallest scales. The blue correlation function is smaller than
than shuffled version, while the red correlation function is
larger. This difference is probably related to the deviations in
Figure \ref{m2_den_color_shuffle} and Figure \ref{bf_den_large_300kpc}
at small scales, and also indicates that either clumps on small scales
within groups have redder galaxy populations, or that our measurements
of projected groupocentric radius are mixing true groupocentric radii.

We have measured the difference between accounting for the
groupocentric distance when shuffling and not doing so. The difference
is negligible on large scales, and minor on small scales.  Thus, at
least for large scale clustering, halo occupation models probably do
not have to account for the radial dependence of halo occupation.

We have explored how these results vary with our choice of color cut
and with galaxy luminosity. If we vary the first term on the right
hand side of Equation \ref{colorcut} between 0.7 and 0.9, the changes
to our results are only 2--3\%. If we use group catalogs including
different ranges of galaxy luminosity ($M_{\band{0.1}{r}} - 5\log_{10}
h<-18$ and $-20$), the results are in Figure \ref{bias}, showing the
relative bias averaged between 4 and 20 $h^{-1}$ Mpc. Both the higher
and lower luminosity samples show more of a relative bias, but the
effects are still smaller than 5\% in each case.

\section{Comparison to previous results}
\label{others}

How do our results compare with similar previous work?
\citet{kauffmann04a} found similar results for the relationship
between D4000 and the large-scale density field at fixed density on
scales of 1 $h^{-1}$ Mpc. \citet{blanton04c} found similar results for
color and H$\alpha$ equivalent width.  \citet{balogh04a} found a
contradictory result that \citet{blanton04c} demonstrated was due to
the sparseness of the catalog used for their density estimators. None
of these tests were as sensitive as those we use here.

Meanwhile, for samples of large clusters \citet{gomez03a} and
\citet{lewis02a} found that the emission line indicators of
star-formation were significantly reduced far outside clusters, around
two to three virial radii. This result suggests that environmental
effects extend very far. As \citet{quintero06a} show using the group
catalog we use here, our estimated virial radii tend to be about twice
theirs, which partly explains the difference in our result. However,
it is also the case that the test in \citet{gomez03a} does not cleanly
separate the effect of small and large-scale environment. The mean
density two to three virial radii away from a large cluster is larger
than the mean density of the universe and tends to be populated by
large groups. As \citet{lewis02a} show, the star-formation rates at
these distances from clusters correlate well with the local density
field (from our point of view here, they correlate with whether
galaxies are in a moderately-sized group or not).  Thus, correlations
of galaxy properties with clustocentric distance, even at large
distances, may still be due to very local physical effects. As we
noted above, in agreement with \citet{lewis02a}, this result indicates
that effects that can only be important in very large clusters such as
ram pressure stripping of the cold disk gas, cannot explain the
relationship between environment and properties observed in less dense
environments.

Recently, \citet{yang06a} have performed a similar test, using the
cross-correlation of groups of fixed mass with galaxies, as a function
of the spectral type of the central group galaxy. For groups with
estimated masses $M> 10^{12}$ $M_\odot$ they find a strong bias on
large scales between the youngest and oldest central galaxies. Their
results indicate that for the central galaxies of large groups the
large-scale density field may indeed be important, while our results
here suggest that the satellites are left relatively affected. In any
case, the changes to the central galaxies do not seem to affect the
correlation function of all galaxies on large scales at more than the
5\% level.

\section{Comparison to theoretical results}
\label{theory}

One assumption certain models of galaxy formation make is that the
mass of a host halo fully determines the statistical properties of the
galaxies within it. That is, they assume that the assembly time of the
halo and its overall merger history are independent of the larger
scale environment. The results of \citet{lemson99a} lent support to
this approximation. However, recent results found correlations to this
effect (\citealt{gao05a}).  \citet{wechsler06a} explored the effect
more fully and found that these correlations were also related to the
concentration of the halos.  They demonstrated that at halos around
$10^{13}$ $M_\odot$ at $z=0$ there is little effect, but at larger
masses, early forming (or high concentration) halos are less
clustered, while at smaller masses early forming halos are more
clustered.

Clearly, our results here mean either: that for the typical masses of
halos in our sample (probably around $10^{12}$--$10^{13}$ $M_\odot$)
the assembly time and other properties are not related to the large
scale density field; or, that the assembly time has little to do with
the whether galaxies in the halo are red or blue. The fact that there
is no dependence even for different ranges of group luminosity (Figure
\ref{bf_den_large_10mpc}) suggests the latter.

\citet{croton06b} have extended the theoretical work to include models
of galaxy formation and follow the consequences of this ``halo
assembly bias.''  For their sample most comparable to ours ($M_{bj}
-5\log_{10} h < -18$) their red galaxy correlation function on large
scales is higher than would be expected based on just the host halo
environment, with a bias of 10\%, while their blue galaxy correlation
function is consistent. While this result is a bit at odds with ours,
we note that there are a number of differences in analysis. For
example, rather than shuffling based on observable properties, they
shuffle based on the predicted halo masses. In addition, they use the
three-dimensional correlation function rather than the projected
correlation function we use here. A fairer comparison would bring
their results fully into observable quantities to test whether our
results here are indeed inconsistent with their hierarchical models.

\section{Conclusions} 
\label{conclusions}

We have described the relationship between galaxy colors and their
group environments, in terms of the luminosity of their host group and
their distance from the center of that group (groupocentric
distance). Furthermore, we have searched for any residual relationship
between galaxy color and measures of the surrounding environment on
various scales, finding only a slight dependence at large scales ($>
1$ $h^{-1}$ Mpc) and a stronger dependence at the smallest scales
($<300$ $h^{-1}$ kpc). Measured at any scale, the variation of galaxy
colors with the surrounding density field does not explain the
variation of color with group environment: the group environment
always yields extra information.

Since other properties, such as concentration and surface brightness,
do not appear to correlate with the density field independently of
color (\citealt{blanton05b,kauffmann04a, quintero06a}) these results
likely extend to other galaxy properties as well. In fact, we have
checked this proposition explicitly for concentration and found no
dependence of concentration on the larger scale density field once the
color and group environment is fixed.

\begin{figure}
\figurenum{\fignum}
\plotone{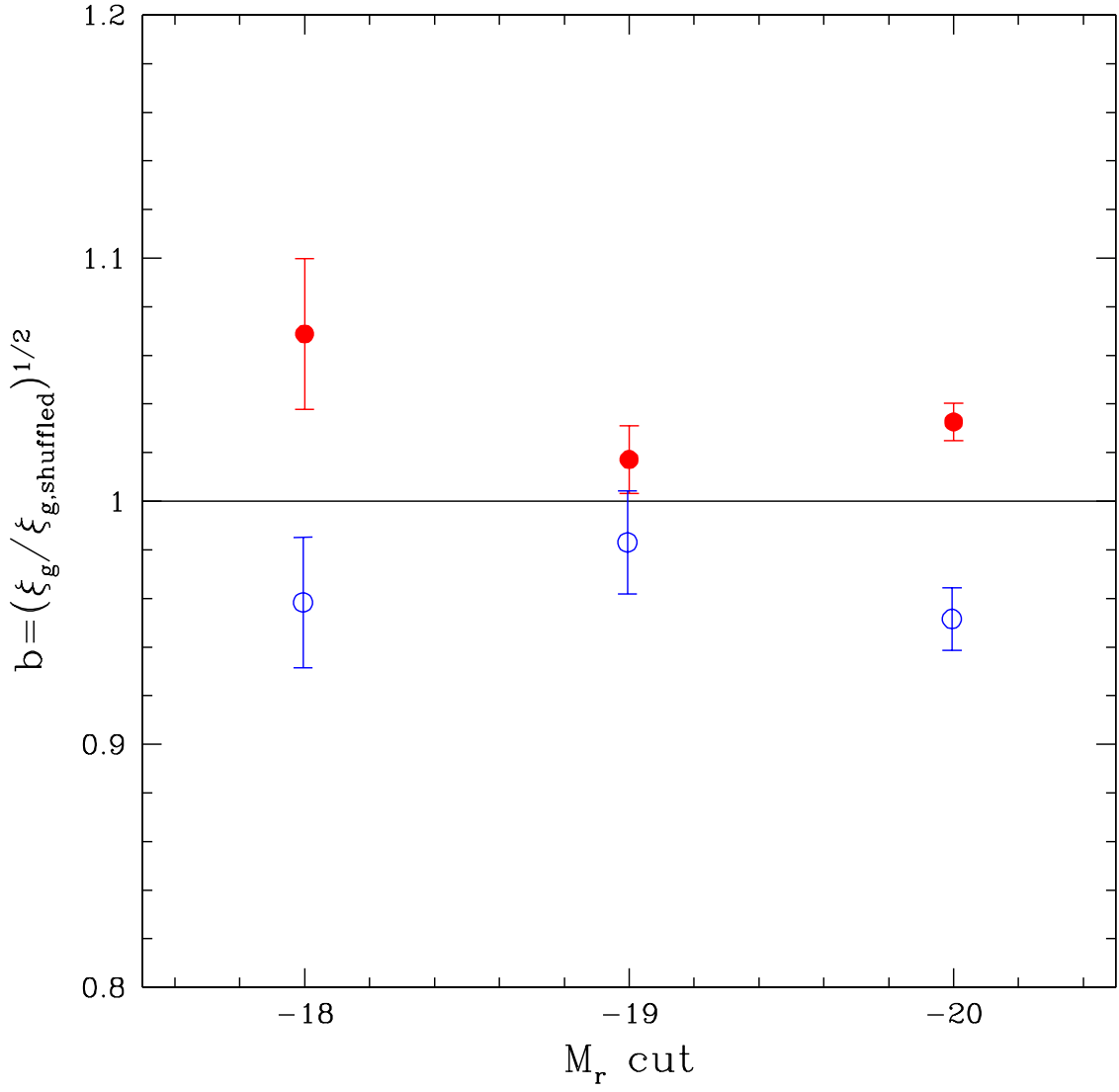}
\caption{\label{bias} Relative bias of the actual galaxy clustering
	relative to the ``shuffled'' galaxy clustering, for three
	volume-limited catalogs. $b$ is averaged between 4 and 20 $h^{-1}$
	Mpc. Solid symbols are the red galaxies, open symbols are the blue
	galaxies.}
\end{figure}

When interpreting these results, keep in mind that we have considered
for the most part only galaxies with $M_r - 5\log_{10} h < -19$
(though in Figure \ref{bias} we also look at slightly lower luminosity
galaxies), which limits us to dark matter halos of $\sim 5\times
10^{11}$ $h^{-1}$ $M_\odot$ or greater. Whether the assembly time or
large-scale density affects galaxies in smaller halos is an open
question.

These results have important consequences for the study of galaxy
formation:
\begin{enumerate}
\item That the blue fraction has no residual relationship with the
	large scale density field (such as 1 $h^{-1}$ Mpc and above)
	demonstrates that galaxy formation is closely tied to physics within
	each halo. The location of that halo in the large-scale density
	field appears not to be important, nor (if CDM is correct) the
	assembly time of the host halos (at fixed halo mass).
\item The residual dependence of color on small scale density may
	indicate the importance of surviving substructure within the group
	(though it is hard to disentangle possible projection effects
	causing this dependence). If so, it may be a signature of the
	processing of blue galaxies into red galaxies in moderately sized
	groups prior to infall into large groups.
\end{enumerate}

These results are also important for studies of large-scale structure
that depend on the use of the halo occupation distribution formalism
to model small-scale clustering. A typical simplification (not a
necessary one) of those models is that how galaxies occupy halos
depends only on their mass, not their larger scale environment. By
showing that the {\it relative} distribution of different types of
galaxies is not affected by the larger scale environment (while it
manifestly is affected by the group luminosity) we lend credence to
this assumption. Furthermore, theoretically speaking, the large-scale
density field is related to the assembly time of the halos, and we
might expect assembly times to be related to the ages of the
galaxies. Either the halos containing our $L_\ast$ galaxy sample do
not have much relationship between their assembly time and large-scale
clustering, or the process of galaxy formation is not much affected by
the assembly time of the halo at a given halo mass. Either possibility
is encouraging to those trying to use the halo model to interpret the
medium-scale correlation functions of similar samples in terms of
cosmological models.

\acknowledgments

We would like to thank Michael Vogeley and Risa Wechsler for useful
comments and suggestions.

Funding for the creation and distribution of the SDSS Archive has been
provided by the Alfred P. Sloan Foundation, the Participating
Institutions, the National Aeronautics and Space Administration, the
National Science Foundation, the U.S. Department of Energy, the
Japanese Monbukagakusho, and the Max Planck Society. The SDSS Web site
is http://www.sdss.org/.

The SDSS is managed by the Astrophysical Research Consortium (ARC) for
the Participating Institutions. The Participating Institutions are The
University of Chicago, Fermilab, the Institute for Advanced Study, the
Japan Participation Group, The Johns Hopkins University, the Korean
Scientist Group, Los Alamos National Laboratory, the
Max-Planck-Institute for Astronomy (MPIA), the Max-Planck-Institute
for Astrophysics (MPA), New Mexico State University, University of
Pittsburgh, University of Portsmouth, Princeton University, the United
States Naval Observatory, and the University of Washington.

\newpage

\clearpage
\clearpage

\setcounter{thefigs}{0}

\clearpage
\stepcounter{thefigs}

\clearpage
\stepcounter{thefigs}

\clearpage
\stepcounter{thefigs}

\clearpage
\stepcounter{thefigs}

\clearpage
\stepcounter{thefigs}

\clearpage
\stepcounter{thefigs}

\clearpage
\stepcounter{thefigs}

\clearpage
\stepcounter{thefigs}

\clearpage
\stepcounter{thefigs}

\clearpage
\stepcounter{thefigs}

\newpage
\clearpage

\end{document}